\documentclass[11pt]{article}
\usepackage{amsmath}
\usepackage{latexsym}
\usepackage{amssymb}
\usepackage{graphicx}
\usepackage{arydshln}

\newtheorem{theorem}{Theorem}[section]
\newtheorem{definition}{Definition}[section]
\newtheorem{lemma}[theorem]{Lemma}
\newtheorem{conjecture}[theorem]{Conjecture}
\newtheorem{corollary}[theorem]{Corollary}

\newtheorem{problem}[theorem]{Problem}
\def\whitebox{{\hbox{\hskip 1pt
 \vrule height 6pt depth 1.5pt
 \lower 1.5pt\vbox to 7.5pt{\hrule width
    3.2pt\vfill\hrule width 3.2pt}%
 \vrule height 6pt depth 1.5pt
 \hskip 1pt } }}
\def\qed{\ifhmode\allowbreak\else\nobreak\fi\hfill\quad\nobreak
     \whitebox\medbreak}

\newcommand{\ignore}[1]{}
\begin{document}

\baselineskip 16pt
\title{ Some new bounds of placement delivery arrays}
\author{\small  X. Niu \ \ and H. Cao \thanks{Research supported by the National Natural
Science Foundation of China under Grant No. 11571179, the natural
science foundation of Jiangsu Province under Grant No. BK20131393,
and the Priority Academic Program Development of Jiangsu Higher
Education Institution. E-mail: caohaitao@njnu.edu.cn
} \\
\small Institue of Mathematics, \\ \small  Nanjing Normal
University, Nanjing 210023, China\\
\small caohaitao@njnu.edu.cn}

\date{}
\maketitle

\begin{abstract}
Coded caching scheme is a technique which reduce  the load during peak traffic times in a wireless network system.
 Placement delivery array (PDA in short) was first introduced by Yan et al.. It can be used to design coded caching scheme.
 In this paper, we prove some lower bounds of PDA on the element and some lower bounds of PDA on the column.
 We also  give some constructions for optimal PDA.

\bigskip

\end{abstract}

\section{Introduction}
Video delivery are becoming the main factor for  the wireless data traffic in the daily life.
With the video on demand and catch-up TV increasing,
the wireless network is predicted to  increase dramatically from 2016 to 2021 \cite{CVN}.
Therefore, communication systems are always slowdown during the peak times.
Caching system is an effective way  to reduce the network slowdown in the peak times,
see \cite{AG,CFL,JCM,KNMD,AN,NM,PMN,STC}.

Maddah-Ali and Niesen propose a centralized coded caching scheme which can
effectively  reduce network slowdown during the peak times.
It use of  network coding theory to caches some contents in the off-peak times.
In a coded caching system, there is a server with $N$ files,
 and each of $K$ users has a cache which is size $M$ and proactively cache $F$, $F\le M$.
 denoted by $(K,F,M,N)$ caching system.
There are two distinct phases in a caching system:

$\bullet$ Placement phase during off-peak times: Parts of content are placed in users¡¯ cache memories.

$\bullet$ Delivery phase during peak times: Requested content is delivered by exploiting the local cache storage
of users.\\

Clearly, it is carried out without the knowledge of the particular user requests in placement phase. And our goal is to minimize
the rate of transmission over the shared link since the delivery phase takes place during peak traffic period. This rate is also
called the{\it delivery rate}.
 Maddah-Ali and Niesen proposed
a deterministic coded caching scheme that utilizes an uncoded combinatorial
cache construction in the placement phase and a linear network code in the delivery phase,
where users store contents in a
coordinated manner in \cite{AN}.
For any size $F$, designing a coded caching scheme with the minimum delivery rate will be a critical issue.
 By investigating placement phase and delivery phase from a combinatorial viewpoint,
a new concept of placement delivery array was introduced to characterize the involved strategies in two phases in \cite{YCTC}.
As a result, the problem of designing  centralized coded caching scheme can be transformed into the problem of designing
a placement delivery array. There are some definitions and results of  placement delivery array in the following.

 We use $[a,b] $ to denote $ \{a,a + 1,...,b\}$ for intervals of integers for any integers $a$ and $b$. If
$a < b$, then we have $[a,b] = \emptyset$. Let $[a,b) = [a,b-1]$, and let $ K,F,S$ be positive integers in this paper.

\begin{definition}
A $(K,F,S)$ placement delivery array ($(K,F,S)-$PDA in short) is an
$F\times K$ array P over $[0,S) \cup \{\ast\}$, if the following two properties are satisfied:
\begin{enumerate}
\item  each integer occurs at most once in each row and in each column of $P$;

 \item whenever two distinct cells $(a,b)$ and $(c,d)$ are occupied by the same integer, the opposite corners $(a,d)$ and $(c,b)$ contain
$\ast$.
\end{enumerate}

 Furthermore, for any positive integer $Z \le F$, $P$ is denoted by $(K,F,Z,S)-$PDA if the symbol $\ast$ appears $Z$ times in each column.
\end{definition}

An $\it{optimal}$ $(K,F,Z,S)-$PDA is a PDA which has the maximum possible $K$
when $F,Z,S$ are given or has the minimum possible $S$ when $K,F,Z$ are given.

\begin{theorem}\label{1}(\cite{CJYTC})
 \begin{enumerate}
\item  For $Z=1,F-1$, an optimal $(K,F,Z,S)-$PDA exists.

\item For $Z=F-2$, $h=\lceil\frac{2K}{F}\rceil$,
 $\lceil\frac{2K}{h+1}\rceil(h+1)\le hF$ and $ (h+1)\mid F$,
 there exists an optimal $(K,F,Z,S)-$PDA except both $h$ and $\frac{F}{h+1}$ are odd.

\item For $Z=F-2$, $h=\lceil\frac{2K}{F}\rceil$, $\lceil\frac{2K}{h+1}\rceil(h+1)> hF$ and $ (h+2)\mid F$,
 there exists an optimal $(K,F,Z,S)-$PDA.
\end{enumerate}
\end{theorem}

\begin{theorem}\label{2}(\cite{AN})
  A $(K,M,N)$ caching system can obtain a $(\binom{F}{t},F,\binom{F}{t-1},\binom{F}{t+1})-$PDA with $t=\frac{KM}{N}$.

\end{theorem}

\begin{theorem}\label{3}
  For positive integers $m,a,b$ satisfying $a+b\le m$, there exists an $(\binom{m}{a},\binom{m}{b},\binom{m}{b}-\binom{m-a}{b},\binom{m}{a+b})-$PDA.
\end{theorem}

\begin{theorem}\label{4}
  For positive integers $m,a,b,c$ with $0< a<m$, $0< b<m$, $0\le c\le $min$\{a,b\}$, there exists a $(K,F,Z,S)-$PDA with $K=\binom{m}{a}, F=\binom{m}{b}, Z=\binom{m}{b}-\binom{m-a}{b-c}\binom{a}{c})$ $S=\binom{m}{a+b-2c}\times$ min$\{ \binom{m-a-b+2c}{c},\binom{a+b-2c}{a-c}\}$.
\end{theorem}

\begin{theorem}\label{r-c-e}
 Let $K,F,Z,S$ be positive integers and $F\ge Z$. If $P_1$ is a $(K,F,Z,S)-$PDA,
 and if $P_2$ is a array obtained from  $P_1$ by permuting the rows and/or columns
and/or elements, then $P_2$ is also a $(K,F,Z,S)-$PDA.
\end{theorem}

This paper is organized as follows. In the next section, we shall give some lower bounds of PDA on $S$.
In Section 3 we will give some properties of PDA  and prove some cases for optimal PDA.
In Section 4 we present a construction for PDA, and use a recursive method to get a lower bound of PDA on $K$.
Section 5 consists of some concluding remarks and directions for future research.

\section{ Lower bounds for $S$ }
Let $S_{(F,F-Z,K)}={\it min} \{S:\ { \it there\ is\ a\ } (K,F,Z,S)-PDA\}$.

Suppose $F,K,Z,$ and $i$ are positive integers. Define $f_{(K,F,Z)}(0)= \lceil\frac{K\cdot(F-Z)}{F}\rceil$, and
$$f_{(K,F,Z)}(i)= \lceil\lceil\lceil\lceil \lceil \frac{K\cdot(F-Z)}{F}\rceil\frac{F-Z-1}{F-1}\rceil \ldots \rceil\frac{F-Z-i+1}{F-i+1} \rceil \frac{F-Z-i}{F-i}\rceil,\ 1\le i\le F-Z-1.$$

We have the first lower bound of PDA on $S$ in the following.
\begin{theorem}\label{YB}
If there exists a $(K,F,Z,S)-$PDA,
then $S \ge \sum\limits_{i=0}^{F-Z-1}{f_{(K,F,Z)}(i)}$.
\end{theorem}

\noindent {\it Proof:} Let $P$ be a $(K,F,Z,S)-$PDA. Form the definition of PDA, there are $(F-Z)K$ cells appearing  integers over $[0,S)$.
By the pigeon hole principle and Theorem \ref{r-c-e},
we have that the first row has at least $\lceil\frac{(F-Z)K}{F}\rceil$ cells appearing distinct integers as follow.
\[
P=\left(\begin{array}{cccc;{2pt/2pt}ccc}
x_1&x_2&\cdots &x_{r_1} &*&\cdots&*  \\
\hdashline[2pt/2pt]
&&&&&&\\
&&&&&&\\
\multicolumn{4}{c;{2pt/2pt}}{\raisebox{2ex}[3pt]{\Large $P_{1}$}}

\end{array}\right)
\]
Thus, we have $r_1\ge \lceil\frac{(F-Z)K}{F}\rceil =f_{(K,F,Z)}(0)$.
Let $M_{0}=\{x_i: 1\le i \le r_1\}$.
So $M_{0}\subset [0,S)$.
Now, we obtain $P_1$  form $P$ by permuting the columns from $r_1+1$ to $K$ and the first row.

It is obvious that the symbol $\ast$ appears $Z$ times in each column of $P_1$.
So  $P_1$  is a $(r_1,F-1,Z,S_1)-$PDA where $S_1= \mid\{x :$ the integer $x$ appears in $P_1\}\mid $.
 By the definition of PDA, we have $M_0\cap \{x :$ the integer $x$ appears in $P_1\}=\emptyset$.

Similar the step above, we have $P_{i}$, $M_{i-1}$, and $\mid M_{i-1}\mid \ge f_{(K,F,Z)}(i-1)$,  $2\le i\le F-Z$.
By the definition of PDA, we have $M_i\cap M_j=\emptyset$ for $0\le i<j \le F-Z-1$.
Thus, we have $S \ge \sum_{i=0}^{F-Z-1}{\mid M_1\mid }\ge \sum_{i=0}^{F-Z-1}f_{(K,F,Z)}(i)$.\qed

\begin{corollary}\label{YB2}
If there exists a $(K,F,F-2,S)-$PDA, then
$S \ge\lceil \frac{2K}{F}\rceil +\lceil \frac{1}{F-1}\lceil \frac{2K}{F}\rceil\rceil.$
\end{corollary}

\begin{theorem}\label{}
If there exists a $(K,F,Z,S)-$PDA, then $S\ge S_{(F-\lceil \frac{(F-Z)\cdot K}{S}\rceil,F-Z-1,\lceil \frac{(F-Z)\cdot K}{S}\rceil)}+1$.
\end{theorem}

\noindent {\it Proof:} Let $P$ be a $(K,F,Z,S)-$PDA. By definition of PDA, there are $(F-Z)K$ cells appearing  integers over $[0,S)$.
By the pigeon hole principle, there exists an element $x$ which appears at least $\lceil\frac{(F-Z)K}{S}\rceil$ times.
Let $P_{1}$ obtain from $P$ by deleting the row which contains element $x$, and the column which does not contain the element $x$.
It is obvious that $P_1$  is a $(\lceil \frac{(F-Z)\cdot K}{S}\rceil,F-\lceil \frac{(F-Z)\cdot K}{S}\rceil,Z+1-\lceil \frac{(F-Z)\cdot K}{S}\rceil,S_1)-$PDA, and $S_1\subset S$ by the definition of PDA. Thus, $S\ge S_{(F-\lceil \frac{(F-Z)\cdot K}{S}\rceil,F-Z-1,\lceil \frac{(F-Z)\cdot K}{S}\rceil)}+1$.\qed

\section{ Optimal PDAs }

Let $K_{(F,F-Z,S)}={\it max}\{K:\ {\it there\ is\ a\ } (K,F,Z,S)-PDA\}$.

It is obvious that we have a upper bound of PDA.
\begin{theorem}\label{YBK}

$K_{(F,F-Z,S)}\le\frac{(Z+1)S} {F-Z} $.

\end{theorem}

\noindent {\it Proof:} By definition of $(K,F,Z,S)-$PDA,
any an element appears at most  $Z+1$ times,
so we have $K_{(F,F-Z,S)}\le\frac{(Z+1)S} {F-Z}$.
If $S=\binom{F}{F-Z-1}$, then we have $ K_{(F,F-Z,S)}= \frac{(Z+1)S} {F-Z}$.\qed

\begin{theorem}\label{K-S}
If there exists a $(K,F,Z,S)-$PDA,
 then there is a $(K,S,S-F+Z,F)-$PDA. Furthermore, $ K_{(F, F-Z, S)}=K_{(S,F-Z,F)}$.
\end{theorem}

\noindent {\it Proof:} Let array $P$ be a $(K,F,Z,S)-$PDA. Then $P=(p_{ij})$
and $i \in [1,F],  j\in [1,K],  p_{ij}\in [1,S]\cup \{*\}$. We denote an array $A=(a_{ij})_{S\times K}$ as following
\[
a_{ij}=\begin{cases}
k             & \text{if } p_{kj}=i,\\
* & \text{ otherwise.}
\end{cases}\]
Now, we only to show that $A$ is a $(K,S,S-F+Z,F)-$PDA.
If $a(i_1,j_1)=a(i_2,j_2)=k$,
then $p(k,j_1)=i_1$ and $p(k,j_2)=i_2$,
so $i_1\neq i_2$, $j_1\neq j_2$ and $a(i_1,j_2)=a(i_2,j_1)=*$.
Otherwise, assume $a(i_1,j_2)=k_1$, then $p(k_1,j_2)=i_1$, a contradiction.
Thus, $A$ is a  $(K,S,S-F+Z,F)-$PDA.

Thus, $ K_{(F, F-Z, S)}\le K_{(S,F-Z,F)}$ and $K_{(S,F-Z,F)}\le  K_{(F, F-Z, S)}$.

So we have $ K_{(F, F-Z, S)}=K_{(S,F-Z,F)}$.\qed

We can use the similar method to prove the following theorem.

\begin{theorem}
If there exists a $(K,F,S)-$PDA, then  there exists a $(a,b,c)-$PDA with $\{a,b,c\}=\{K,F,S\}$.

\end{theorem}

Now, we will  consider the $S\ge F$ case in this paper.

\begin{lemma}\label{S_1+S_2}
$K_{(F,F-Z,S_1+S_2)} \ge K_{(F,F-Z,S_1)} +K_{(F,F-Z,S_2)}$.
\end{lemma}

\noindent {\it Proof:} To prove the lemma, we use the  construction.
Let $P_1$  be a $( K_{(F,F-Z,S_1)},F,Z,S_1)-$PDA appearing  integers over $[0,S_1)$,
 and $P_2$ be a $( K_{(F,F-Z,S_2)},F,Z,S_2)-$PDA appearing  integers over $[S_1,S_1+S_2)$.
 Let $P$ be an array as follow.

 \[
P=\left(\begin{array}{c;{2pt/2pt}c}
 P_1    &P_{2}
\end{array}\right)
\]

It is easy to check that each integer occurs at most once in each row and in each column of $P$,
and if two distinct cells $(a,b)$ and $(c,d)$ are occupied by the same integer,
then the two cells are contained in the same array $P_i, 1\le i\le 2$.
So the opposite corners $(a,d)$ and $(c,b)$ contain
$\ast$, by the definition. Thus, $P$ is a $( K_{(F,F-Z,S_1)}+ K_{(F,F-Z,S_2)},F,Z,S_1+S_2)-$PDA.

So we have $K_{(F,F-Z,S_1+S_2)} \ge K_{(F,F-Z,S_1)} +K_{(F,F-Z,S_2)}$.\qed

By Theorem \ref{2}, we have the following theorem.

\begin{theorem}\label{mS}
$K_{(F,F-Z,m\binom{F}{Z+1})} =m\cdot K_{(F,F-Z,\binom{F}{Z+1})}=m\binom{F}{Z}, m\ge 1$.
\end{theorem}

\noindent {\it Proof:}   Let $P_0$ be  a $(\binom{F}{Z},F,Z,\binom{F}{Z+1})-$PDA.
It is obvious that $P_0$ is optimal, otherwise,  suppose that there exists
a $(\binom{F}{Z}+1,F,Z,\binom{F}{Z+1})-$PDA.
By the pigeon hole principle, there exists an element $x$
which appears at least $\lceil\frac{(\binom{F}{Z}+1)(F-Z)}{\binom{F}{Z+1}}\rceil= Z+2$ times in array $P_0$.
By the definition of PDA, the element $x$ appear at least $Z+2$  rows, so the column which contains the element $x$, at least contains $Z+1$
$*$. It contradicts with the definition.
We obtain an array $P_i, 1\le i\le m-1$ from $P_0$ by permuting  the element $j$ to $i\binom{F}{Z+1}+j$, $0\le j< \binom{F}{Z+1}$,
then we have the arrays $P_1,P_2,\cdots,P_{m-1}$. By Theorem \ref{r-c-e},
the arrays $P_1,P_2,\cdots,P_{m-1}$ are  $(\binom{F}{Z},F,Z,\binom{F}{Z+1})-$PDA.
 We construct the array $P$ as follow.
\[
P=\left(\begin{array}{c;{2pt/2pt}c;{2pt/2pt}c;{2pt/2pt}c;{2pt/2pt}c}
P_0     & P_1    &P_{2}  & \cdots  & P_{m-1}
\end{array}\right)
\]
It is easy to check that $P$ is a $(m\binom{F}{Z},F,Z,m\binom{F}{Z+1})-$PDA.
 Thus, $K_{(F,F-Z,m\binom{F}{Z+1})}\ge m\cdot K_{(F,F-Z,\binom{F}{Z+1})}=m\binom{F}{Z}, m\ge 1$.

To prove $K_{(F,F-Z,m\binom{F}{Z+1})}\le m\cdot K_{(F,F-Z,\binom{F}{Z+1})}=m\binom{F}{Z}, m\ge 1$,
we assume that there exists a $(m\binom{F}{Z}+1,F,Z,m\binom{F}{Z+1})-$PDA and get a contradiction.

Assume that $P$ is a $(m\binom{F}{Z}+1,F,Z,m\binom{F}{Z+1})-$PDA.
From the definition of PDA, there are $(F-Z)(m\binom{F}{Z}+1)$ cells appearing  integers over $[0,m\binom{F}{Z+1})$.
By the pigeon hole principle, there exists an element $y$
which appears at least $\lceil\frac{(m\binom{F}{Z}+1)(F-Z)}{m\binom{F}{Z+1}}\rceil= Z+2$ times in array $P$.
By the definition of PDA, the element $y$ appears at least $Z+2$  rows, so the column which contains the element $y$, at least contains $Z+1$
$*$. It contradicts with the define.

Thus, $K_{(F,F-Z,m\binom{F}{Z+1})} =m\cdot K_{(F,F-Z,\binom{F}{Z+1})}=m\binom{F}{Z}, m\ge 1$.\qed

\begin{lemma}\label{2-S}
$K_{(2,2,S)}=\frac{S-1}{2}+\frac{d-1}{2}$, where $d=gcd(2,S)$.
\end{lemma}

\noindent {\it Proof:} By  Lemma \ref{S_1+S_2},
we have $K_{(2,2,S)}\ge\frac{S-1}{2}+\frac{d-1}{2}$, where $d=gcd(2,S)$.
Assume that $K_{(2,2,S)}=\frac{S-1}{2}+\frac{d-1}{2}+1=\frac{S+d}{2}$.
Let $P$ be a $(\frac{S+d}{2},2,0,S)-$PDA. By the pigeon hole principle, there exists an element $x$
which appears at least $\lceil\frac{S+d}{S}\rceil= 2$ times in array $P$, a contradiction.\qed

From the properties of PDA, we give a problem as follow.
\begin{problem}\label{P1}
 Let  $d=gcd(F,S)$, prove $K_{(F,2,S)}=\frac{(F-1)(S-1)}{2}+\frac{d-1}{2}$.
\end{problem}

There are some cases that the Problem \ref{P1} is right.

\begin{theorem}\label{OPr}
Let $m_1,m_2,F$ be positive  integers. Then
 $K_{(F,2,m_1\cdot F+ r)} =m_1\cdot K_{(F,2,F)} +m_2\cdot K_{(r,2,r)}$, if $F=r\cdot m_2$.
\end{theorem}

\noindent {\it Proof:} By Theorems \ref{K-S} and \ref{mS}, Lemma \ref{S_1+S_2},
we have $K_{(F,2,m_1\cdot F+ r)} \ge m_1\cdot K_{(F,2,F)} +m_2\cdot K_{(r,2,r)}=\frac{m_1F(F-1)+m_2r(r-1)}{2}$.

Suppose that there exists an array $P$ which is a $(\frac{m_1F(F-1)+m_2r(r-1)}{2}+1,F,F-2,m_1F+r)-$PDA,
then there are $m_1F(F-1)+m_2r(r-1)+2$ cells appearing integer over $[0,m_1F+r)$.
By the pigeon hole principle, there exists a row $l$
which  appears at least $\lceil \frac{m_1F(F-1)+m_2r(r-1)+2}{F}\rceil=m_1(F-1)+r$  elements.
So the array $P$ contains at least $\lceil \frac{m_1(F-1)+r}{F-1}\rceil=m_1+1$ distinct elements which are different to the elements in row $l$.
Thus, $m_1F+r \ge m_1(F-1)+r+m_1+1=m_1F+r+1 $, a contradiction.
Thus, $K_{(F,2,m_1\cdot F+ r)} =m_1\cdot K_{(F,2,F)} +m_2\cdot K_{(r,2,r)}$, if $F=r\cdot m_2$. \qed

\begin{theorem}\label{OP1}
Let $F\ge 2$ be positive  integers. Then
\begin{enumerate}
\item $K_{(F,2,mF+1)} = mK_{(F,2,F)}$;

\item $K_{(F,2,mF+2)} =mK_{(F,2,F)} +K_{(F,2,2)}$.
\end{enumerate}
\end{theorem}

\noindent {\it Proof:} 1. By Lemma  \ref{S_1+S_2} and Theorem \ref{mS},
we have $K_{(F,2,m\cdot F+ 1)} \ge m\cdot K_{(F,2,F)} =\frac{mF(F-1)}{2}$.

Suppose that there exists an array $P$ which is a $(\frac{mF(F-1)}{2}+1,F,F-2,mF+1)-$PDA,
then there are $mF(F-1)+2$ cells appearing integer over $[0,mF+1)$.
By the pigeon hole principle, there exists a row $l$
which  appears at least $\lceil \frac{mF(F-1)+2}{F}\rceil=m(F-1)+1$  elements.
So the array $P$ contain at least $\lceil \frac{m(F-1)+1}{F-1}\rceil=m+1$ distinct elements which are different to the elements in row $l$.
So, $mF+1 \ge m(F-1)+1+m+1=mF+2$, a contradiction.
Thus, $K_{(F,2,m\cdot F+ 1)} =m\cdot K_{(F,2,F)} $.

 2. By Theorems  \ref{K-S}, \ref{mS}  and  Lemmas  \ref{S_1+S_2}, \ref{2-S},
we have $K_{(F,2,m\cdot F+ 2)} \ge m\cdot K_{(F,2,F)}+K_{(2,2,F)} =\frac{mF(F-1)}{2}+\lfloor\frac{F}{2}\rfloor$.

Suppose that there exists an array $P$ which is a $(\frac{mF(F-1)}{2}+\lfloor\frac{F}{2}\rfloor+1,F,F-2,mF+2)-$PDA,
then there are at least $mF(F-1)+F+1$ cells appearing integer over $[0,mF+2)$.
By the pigeon hole principle, there exists a row $l$
which  appears at least $\lceil \frac{mF(F-1)+F+1}{F}\rceil=m(F-1)+2$  elements.
So the array $P$ contains at least $\lceil \frac{m(F-1)+2}{F-1}\rceil=m+1$ distinct elements which are different to the elements in row $l$.
So, $mF+2 \ge m(F-1)+2+m+1=mF+3$, a contradiction.
Thus, $K_{(F,2,m\cdot F+ 2)} =m\cdot K_{(F,2,F)}+K_{(2,2,F)} $.\qed

\begin{theorem}\label{OPF-1}
Let $F\ge4$ be positive  integers. Then
\begin{enumerate}
\item $K_{(F,2,mF+F-2)} = mK_{(F,2,F)} +K_{(F-2,2,F-2)}+ K_{(2,2,F-2)}$;

\item $K_{ (F,2,mF+F-1)} = mK_{(F,2,F)} +K_{(F-1,2,F-1)}$.
\end{enumerate}
\end{theorem}

\noindent {\it Proof:}  1. By Theorems  \ref{K-S}, \ref{mS}  and  Lemma \ref{S_1+S_2},
we have $K_{(F,2,mF+F-2)} \ge mK_{(F,2,F)} +K_{(F-2,2,F-2)}+ K_{(2,2,F-2)}=\frac{mF(F-1)+(F-2)(F-3)}{2}+\lfloor\frac{F-2}{2}\rfloor$.

Suppose that there exists  a $(\frac{mF(F-1)+(F-2)(F-3)}{2}+\lfloor\frac{F-2}{2}\rfloor+1,F,F-2,mF+F-2)-$PDA.
By Theorem \ref{PJD}, we have $mF+F-2 \ge\lceil\frac{mF(F-1)+(F-2)(F-3)+F+1)+(2-F)(mF+F-2)}{F}\rceil F=mF +F$, a contradiction.
Thus, $K_{(F,2,m\cdot F+ F-2)} =m\cdot K_{(F,2,F)} +K_{(F-2,2,F-2)}+ K_{(2,2,F-2)}$.

 2. By Theorems  \ref{K-S}, \ref{mS}  and  Lemma \ref{S_1+S_2},
we have $K_{(F,2,m\cdot F+ F-1)} \ge m\cdot K_{(F,2,F)}+K_{(F-1,2,F-1)} =\frac{mF(F-1)}{2}+\frac{(F-1)(F-2)}{2}$
$=\frac{mF(F-1)+(F-1)(F-2)}{2}$.

Suppose that there exists a $(\frac{mF(F-1)+(F-1)(F-2)}{2}+1,F,F-2,mF+F-1)-$PDA.
By Theorem \ref{PJD}, we have $mF+F-1 \ge\lceil\frac{mF(F-1)+(F-1)(F-2)+2)+(2-F)(mF+F-1)}{F}\rceil F=mF +F$, a contradiction.
Thus, $K_{(F,2,m\cdot F+ F-1)} =m\cdot K_{(F,2,F)} +K_{(F-1,2,F-1)}$.\qed

\begin{theorem}
$K_{(F,2,S)}=\frac{(F-1)(S-1)}{2}+\frac{d-1}{2}$, where $F\in \{2,3,4,5,6\}$ and $d=gcd(F,S)$.
\end{theorem}
\noindent {\it Proof:}
It is obvious that the theorem hold by Theorems  \ref{OPr}, \ref{OP1} and \ref{OPF-1}.\qed

In order to prove Problem \ref{P1}, we only need to consider the following problem.

\begin{problem}\label{}
 Let $r$ be an positive integer. For $F\ge 7$ and $3\le r\le F-3$, prove
$K_{(F,2,F+r)} = K_{(F,2,F)} +K_{(F,2,r)}$.
\end{problem}

\section{Lower bounds for $K$}

In this section, let  $F\ge 7 $ and  $S=mF+r$, $3\le r\le F-3$,
$m\ge 1$ be positive integers, and let $d=gcd(F,S)$. Thus, $d=gcd(F,r)$.

\noindent Let $d(x)$ be the element $x$ appearing times in the PDA.

\noindent Let $A_{x}=\{i \ | \ 1\le i\le F, \it{ the ~element} \ x ~does ~not ~appear ~in ~the ~i$-$th ~row\}$.

If $d(x)=F-1$, then $\mid A_x\mid=1$.

\begin{lemma}\label{2BD}
$K_{(F,2,S)}\ge\frac{(F-1)\cdot (S-1)}{2}+\frac{d-1}{2}$, where $d=gcd(F,S)$.
\end{lemma}

\noindent {\it Proof:} We use induction on $F$ to prove this lemma.

1. By Lemma~\ref{2-S}, the case $F=2$  holds.

2.  Suppose that for  $F<k, k\ge 3$,
we  have $K_{(F,2,S)}\ge\frac{(F-1)\cdot (S-1)}{2}+\frac{d-1}{2}$.

3. Let $F=k$, and $S=mF+r, 1\le r\le F$. Since $d=gcd(F,S)$, then  we have $d=gcd(F,r)$.
By Theorems  \ref{K-S}, \ref{mS}  and  Lemma \ref{S_1+S_2},
   we have

$K_{(F,2,S)}\ge K_{(F,2,mF)}+K_{(F,2,r)}$

\hspace*{1.3cm}$=\frac{mF(F-1)}{2} +K_{(r,2,F)}$

\hspace*{1.3cm}$\ge\frac{mF(F-1)}{2} +\frac{(r-1)\cdot (F-1)}{2}+\frac{d-1}{2}$

\hspace*{1.3cm}$=\frac{(mF+r-1)(F-1)}{2} +\frac{d-1}{2}$

\hspace*{1.3cm}$=\frac{(F-1)\cdot (S-1)}{2}+\frac{d-1}{2}$\qed

\begin{lemma}\label{MAXD}
If there exists a $(K_{(F,2,S)},F,F-2,S)-$PDA, then each element appear at most  $ \lceil\frac{2K_{(F,2,S)}}{S}\rceil=F-1$ times.
Furthermore, there are at least $S-(F-d)$ elements which appear $F-1$ times, where $d=gcd(F,S)$.
\end{lemma}

\noindent {\it Proof:} By Lemma \ref{2BD},
we have $K_{(F,2,S)}\ge \frac{(F-1)(S-1)+d-1}{2}$ where  $d=gcd(F,S)$.

So $\lceil\frac{2K_{(F,2,S)}}{S}\rceil\ge \lceil\frac{(F-1)(S-1)+d-1}{S}\rceil$

\hspace*{2.27cm} $=\lceil\frac{(F-1)S -F +d}{S}\rceil$

\hspace*{2.27cm} $=\lceil\frac{(F-2)S+S-F +d}{S}\rceil$

\hspace*{2.27cm} $=\lceil F-2+\frac{S-F+d}{S}\rceil$

\hspace*{2.27cm} $= F-1$.

By the definition of PDA, we have an element appear at most $F-1$ times.
 Thus, there are  at least $S-(F-d)$ elements which appear $F-1$ times.\qed

\begin{theorem}\label{PJD}
 If there exists a $(K_{(F,2,S)},F,F-2,S)-$PDA, then
$$S \ge\lceil \frac{2K_{(F,2,S)}+2S-SF}{F}\rceil F.$$
\end{theorem}

\noindent {\it Proof:} Let $P$ be a $(K_{(F,2,S)},F,F-2,S)-$PDA. From the definition of PDA, we have

$2K_{(F,2,S)}=\sum\limits_{x\in[0,S)}{d(x)}$
$=\sum\limits_{\substack{x\in[0,S)\\d(x)\ge \lceil\frac{2K_{(F,2,S)}}{S}\rceil}}  {d(x)}
+\sum\limits_{\substack{x\in[0,S)\\d(x)< \lceil\frac{2K_{(F,2,S)}}{S}\rceil}}  {d(x)}$

\hspace*{4cm}$\le\sum\limits_{\substack{x\in[0,S)\\d(x)=F-1}}  {(F-1)}
+\sum\limits_{\substack{x\in[0,S)\\d(x)< F-1}}  {(F-2)}$.

So $\sum\limits_{\substack{x\in[0,S)\\d(x)=F-1}}  {1}\ge 2K_{(F,2,S)}-S(F-2) $.

Since $d(x)=F-1$, then there has  only one row which not contains the element $x$ in array $P$.
 By the pigeon hole principle, there exists a row $l$
which not appears at least $\lceil \frac{2K_{(F,2,S)}+2S-SF}{F}\rceil$  elements in array $P$.
So there are at least $\lceil \frac{2K_{(F,2,S)}+2S-SF}{F}\rceil(F-1)$ distinct elements in the row $l$.
Thus, $S \ge\lceil \frac{2K_{(F,2,S)}+2S-SF}{F}\rceil F$.\qed

\begin{lemma}\label{MAXE}
If there exists a $(K_{(F,2,S)},F,F-2,S)-$PDA, and let $S=mF+r$ and $d=gcd(F,S)$,
then each row contains at most $ \lceil\frac{2K_{(F,2,S)}}{F}\rceil=S-(m+1)$ element.
Furthermore,  there are at least $F-r+d$ rows which contain $S-(m+1)$ elements.
\end{lemma}
\noindent {\it Proof:} By Lemmas \ref{S_1+S_2} and \ref{2BD},
we have $K_{(F,2,S)}\ge K_{(F,2,mF)} +K_{(F,2,r)} \ge \frac{mF(F-1)+(F-1)(r-1)+d-1}{2}$.

So $\lceil\frac{2K_{(F,2,S)}}{F}\rceil\ge \lceil\frac{(F-1)(mF+r-1)+d-1}{F}\rceil$

\hspace*{2.23cm} $=m(F-1)+\lceil\frac{(F-1)(r-1)+d-1}{F}\rceil$

\hspace*{2.23cm} $=m(F-1)+\lceil\frac{F(r-2)+F-r+d}{F}\rceil$

\hspace*{2.23cm} $=m(F-1)+(r-2)+\lceil\frac{F-r+d}{F}\rceil$

\hspace*{2.23cm} $=m(F-1)+(r-2)+1$

\hspace*{2.23cm} $=mF+r-m-1$

\hspace*{2.23cm} $=S-(m+1)$.

Suppose that there exists one row which contain $S-m$ elements.
Without loss of generality, we can assume the first row satisfying the hypothesis.
By Corollary \ref{YB2}, we have $S\ge S-m+\lceil\frac{mF+r-F}{F-1}\rceil= S-m+ m+1=S+1$,
a contradiction. So each row contain at most $S-(m+1)$ elements.
Thus, there are at least $F-r+d$  rows where each row contains $S-(m+1)$ elements.\qed

\begin{lemma}\label{NAR}
If there exists a $(K_{(F,2,S)},F,F-2,S)-$PDA with $m > F-r-d$,
then there exist $F$ elements $x_1, x_2, \ldots, x_F$ with $d(x_i)=F-1, 1\le i\le F$
such that $A_{x_i}=\{i\}$.
\end{lemma}

\noindent {\it Proof:} Assume that $ \mid\bigcup\limits_{\substack{x\in[0,S)\\d(x)=F-1}}{A_{x}}  \mid \le F-1$.
 By Lemma \ref{MAXE}, there are at least $F-r+d$
 rows  which contain $S-(m+1)$ elements.

 By Lemma \ref{MAXD}, there are at least $mF+r-(F-d)$
 elements which repeat appearing $F-1$ times.
By the pigeon hole principle, $\lceil\frac{mF+r-(F-d)}{F}\rceil=m-1 +\lceil\frac{r+d}{F}\rceil= m$.
If there exists a row which does not contain $m+1$ elements  appearing $F-1$ times,
 then this row contains at least $(m+1)(F-1)$ distinct elements.
 So $S \ge (m+1)(F-1)+m+1=S +(F-r)$, a contradiction.
  Thus, there are at least $r+d$  rows which
 not contain $m$ elements which appear $F-1$ times.
Since $F-r+d+(r+d)=F+ 2d$, then we have at least $2d$ rows which
 contain $S-(m+1)$ elements and does not contain $m$ elements  appearing $F-1$ times.
By Theorem \ref{r-c-e},  the first row satisfies the two properties
and  contains $S-(m+1)$ elements in the first $S-(m+1)$ columns.
So the first $S-(m+1)$ columns contain $S$ distinct elements.

 Since $ \mid\bigcup\limits_{\substack{x\in[0,S)\\d(x)=F-1}}{A_{x}}  \mid \le F-1$,
 then there exists a row not appearing $\lceil\frac{mF+r-(F-d)-m}{F-2}\rceil = \lceil m
 +\frac{m-F+r+d}{F-2}\rceil=m+1$ elements which repeat appearing $F-1$, a contradiction.
Thus,  we have  $F$ elements $x_1, x_2, \ldots, x_F$ with $d(x_i)=F-1, 1\le i\le F$
such that $A_{x_i}=\{i\}$.\qed

\begin{lemma}\label{SA1}
Let $P$ be  a $(K_{(F,2,S)},F,F-2,S)-$PDA with $m> F-r-d$.
Then there exists a subarray of a $(K_{(F,2,S)},F,F-2,S)-$PDA as following,
{\tiny \[
P_1=\left(\begin{array}{ccccc;{2pt/2pt}c;{2pt/2pt}c;{2pt/2pt}c;{2pt/2pt}c;{2pt/2pt}c;{2pt/2pt}c}
x_1&x_2&\cdots &x_{F-2} &x_{F-1}  & \overbrace{\cdots}^{F-1} & \overbrace{\cdots}^{F-1}& \cdots & \overbrace{\cdots}^{F-1} & \overbrace{\cdots}^{r-1}& \overbrace{**\cdots*}^{K_{(F,2,S)}-S+ m+1}\\
x_{F}&*&\cdots&*&*&\cdots&\cdots&\cdots&\cdots&\cdots&\overbrace{\cdots}^{K_{(F,2,S)}-S+ m+1}\\
*& x_{F}&\cdots&*&*&\cdots&\cdots&\cdots&\cdots&\cdots&\cdots\\
\vdots&\vdots&\ddots&\vdots&\vdots&\vdots&\vdots&\vdots&\vdots&\vdots&\vdots\\
*&*&\cdots&x_{F}&*&\cdots&\cdots&\cdots&\cdots&\cdots&\cdots\\
*&*&\cdots&*&x_{F}&\cdots&\cdots&\cdots&\cdots&\cdots&\cdots
\end{array}\right)_{F\times K_{(F,2,S)}}
\]}
and $d(x_{i})=F-1, 1\le i\le F$.
\end{lemma}

\noindent {\it Proof:}  By Theorem \ref{r-c-e} and Lemma \ref{MAXD}, we have a subarray
{\tiny\[
P_2=\left(\begin{array}{ccccc;{2pt/2pt}c;{2pt/2pt}c;{2pt/2pt}c;{2pt/2pt}c;{2pt/2pt}c;{2pt/2pt}c}
1&2&\cdots & F-2 &F-1  & \overbrace{\cdots}^{F-1} & \overbrace{\cdots}^{F-1}& \cdots & \overbrace{\cdots}^{F-1} & \overbrace{\cdots}^{r-1}&\overbrace{**\cdots*}^{K_{(F,2,S)}-S+ m+1}\\
F&*&\cdots&*&*&\cdots&\cdots&\cdots&\cdots&\cdots&\overbrace{\cdots}^{K_{(F,2,S)}-S+ m+1}\\
*& F&\cdots&*&*&\cdots&\cdots&\cdots&\cdots&\cdots&\cdots\\
\vdots&\vdots&\ddots&\vdots&\vdots&\vdots&\vdots&\vdots&\vdots&\vdots&\vdots\\
*&*&\cdots&F&*&\cdots&\cdots&\cdots&\cdots&\cdots&\cdots\\
*&*&\cdots&*&F&\cdots&\cdots&\cdots&\cdots&\cdots&\cdots
\end{array}\right)_{F\times K_{(F,2,S)}}
\]}
where $d(F)=F-1$.

If $d(i)=F-1, 1\le i\le F-1$, then we let $x_{i}=i$. So we obtain the subarray $P_1$;
If $d(i)<F-1, 1\le i\le F-1$, then there exists an elements $x$ such that $A_{x}=\{i+1\}$,
then $d(x)=F-1 $, by Lemma \ref{NAR}.
So the first row contains the element $x$ and let $p_{1,l}=x$.
Since $d(x)=F-1$, then $p_{k,l}=*$ where $k \neq 1$ and $k \neq i+1$, by the definition of PDA.
we obtain an array $P_{1}^{\prime}$ from $P_{1}$ by letting $p_{1,i}=x$ and $p_{1,l}=i$.
 Now, we show that $P_{1}^{\prime}$ satisfies the definition of PDA.
  Since $P_1$ is a PDA, then we have that $P_{1}^{\prime}$ satisfies the property 1,
 and if $p_{a,b}=p_{c,d}\notin \{i, x\}$, then the opposite corners $p_{a,d}=p_{c,b}=*$, and
 if $p_{a,b}=p_{c,d} =i$ or $x$ and $1 \neq \{a,c\}$, then the opposite corners $p_{a,d}=p_{c,b}=*$.

If  $p_{1,i}=p_{a,b} =x$, then $p_{a,i}=p_{1,b}=*$, otherwise $p_{a,i}\neq *$ or $p_{1,b}\neq*$.
 Thus, $p_{a,i}\neq *$ or $p_{1,b}\neq*$  in array $P_1$.
 In the array $P_{1}$, $p_{a,i}=*$, if $p_{1,b}\neq*$, then it contradicts with the $p_{a,b}=p_{1,l}=x$;

If  $p_{1,l}=p_{c,d} =i$, then $p_{c,l}=p_{1,d}=*$, otherwise $p_{1,d}\neq *$.
 Thus, $p_{1,d}\neq *$  in array $P_1$.
 In the array $P_{1}$, if $p_{1,d}\neq*$, then it contradicts with the $p_{c,d}=p_{1,i}=x$;

 Thus, $P_{1}^{\prime}$ satisfies the  definition of PDA. So  we obtain  the  array $P_1$ from $P_{1}^{\prime}$ by letting $x_{j}=p_{1,j}, 1\le j\le F-1$ and $x_{F}=F$. \qed

\begin{lemma}\label{SA2}
Let $P$ be  a $(K_{(F,2,S)},F,F-2,S)-$PDA with $m> F-r-d$.
Then there exists a subarray of a $(K_{(F,2,S)},F,F-2,S)-$PDA as following,
{\tiny\[
P_0=\left(\begin{array}{ccccc;{2pt/2pt}cccc;{2pt/2pt}ccc;{2pt/2pt}c;{2pt/2pt}c}
x_1    & x_2    & \cdots  & x_{F-2} & x_{F-1}& *     &  *     & \cdots   &*       &*   &\cdots    &\ast    &\cdots& *\\
x_{F}  & \ast   & \cdots  & \ast    & \ast   & x_2   & x_{3}  & \cdots   & x_{F-1}&*   &\cdots    & *      &\cdots& *\\
\ast   &  x_{F} & \cdots  & \ast    & \ast   &x_1    & \ast   & \cdots   & \ast  &x_3  & \cdots   & x_{F-1}&\cdots&\vdots\\
\vdots & \vdots & \ddots  &\vdots   & \vdots &\ast   &  x_1   & \cdots   & \ast  &x_2  &\cdots    &\ast    &\cdots& *\\
\ast   & \ast   &\cdots   &x_{F}    &  \ast  & \vdots& \vdots & \ddots   & \vdots&\vdots & \ddots & \vdots &\cdots& x_{F-1}\\
\ast   &  \ast  &\cdots   &  \ast   &  x_{F} &\ast   &  \ast  &\cdots    &  x_1  &*    &\cdots    & x_2    &\cdots& x_{F-2}\\
\end{array}\right)_{F\times \frac{F(F-1)}{2}}.
\]}
\end{lemma}

\noindent {\it Proof:} By Lemma \ref{SA1}, we have an array $P_1$
which $d(x_{i})=F-1, 1\le i\le F$, $p_{1,j}\neq *, 1\le j\le S-(m+1)$
and $p_{1,j}= *, j> S-(m+1)$ as following.
{\tiny \[
P_1=\left(\begin{array}{ccccc;{2pt/2pt}c;{2pt/2pt}c;{2pt/2pt}c;{2pt/2pt}c;{2pt/2pt}c;{2pt/2pt}c}
x_1&x_2&\cdots &x_{F-2} &x_{F-1}  & \overbrace{\cdots}^{F-1} & \overbrace{\cdots}^{F-1}& \cdots & \overbrace{\cdots}^{F-1} & \overbrace{\cdots}^{r-1}& \overbrace{**\cdots*}^{K_{(F,2,S)}-S+ m+1}\\
x_{F}&*&\cdots&*&*&\cdots&\cdots&\cdots&\cdots&\cdots\\
*& x_{F}&\cdots&*&*&\cdots&\cdots&\cdots&\cdots&\cdots\\
\vdots&\vdots&\ddots&\vdots&\vdots&\vdots&\vdots&\vdots&\vdots&\vdots\\
*&*&\cdots&x_{F}&*&\cdots&\cdots&\cdots&\cdots&\cdots\\
*&*&\cdots&*&x_{F}&\cdots&\cdots&\cdots&\cdots&\cdots
\end{array}\right)_{F\times K_{(F,2,S)}}
\]}
By  the definition of PDA,
the columns from $F$ to $S-m-1$ does not contain the elements $x_{i}, 1\le i\le F-1$.
So we obtain an array  $P_1^{\prime}$ from $P_1$ by deleting the first $S-m-1$ columns.
Then there are all $*$ in the first row of $P_1^{\prime}$.
So we consider the rows from $2$ to $F$, the elements
$x_{i}, 1\le i\le F-1$ appear $F-2$ times. By Lemma \ref{SA1}, we obtain a subarray $P_2$ from $P_1^{\prime}$, as following,
{\tiny \[
P_2=\left(\begin{array}{ccccc;{2pt/2pt}c;{2pt/2pt}c}
*&*&\cdots&*&* &\overbrace{**\cdots*}& \overbrace{**\cdots*}\\
x_2&x_3&\cdots &x_{F-2} &x_{F-1} &\cdots & \overbrace{**\cdots*} \\
x_{1}&*&\cdots&*&*\\
*& x_{1}&\cdots&*&*\\
\vdots&\vdots&\ddots&\vdots&\vdots\\
*&*&\cdots&x_{1}&*\\
*&*&\cdots&*&x_{1}
\end{array}\right)
\]}
Then repeat the step above, we obtain the array $P_k, 3\le k\le F-1$  as following,
{\tiny \[
P_k=\left(\begin{array}{ccccc;{2pt/2pt}c;{2pt/2pt}c}
*&*&\cdots&*&* &\overbrace{**\cdots*}& \overbrace{**\cdots*}\\
\vdots&\vdots&\ddots&\vdots&\vdots&\vdots&\vdots\\
*&*&\cdots&*&* &\overbrace{**\cdots*}& \overbrace{**\cdots*}\\
x_k&x_{k+1}&\cdots &x_{F-2} &x_{F-1} &\cdots & \overbrace{**\cdots*} \\
x_{k-1}&*&\cdots&*&*\\
*& x_{k-1}&\cdots&*&*\\
\vdots&\vdots&\ddots&\vdots&\vdots\\
*&*&\cdots&x_{k-1}&*\\
*&*&\cdots&*&x_{k-1}
\end{array}\right)
\]}

By Theorem \ref{r-c-e}, we can obtain the array $P_0$ from $P$.\qed

\begin{theorem}\label{}

$K_{(F,2,mF+r)} = K_{(F,2,F)} +K_{(F,2,(m-1)F+r)}= \frac{F(F-1)}{2}+K_{(F,2,(m-1)F+r)}$ where $m> F-r-d$.
\end{theorem}

\noindent {\it Proof:} 1. By Lemma \ref{S_1+S_2} and Theorem \ref{mS},
we have $K_{(F,2,mF+r)} \ge K_{(F,2,F)} +K_{(F,2,(m-1)F+r)}= \frac{F(F-1)}{2}+K_{(F,2,(m-1)F+r)}$.

2. We will show that $K_{(F,2,mF+r)} \le K_{(F,2,F)} +K_{(F,2,(m-1)F+r)}$,
i.e.$K_{(F,2,(m-1)F+r)}\ge K_{(F,2,mF+r)} - K_{(F,2,F)}$.

Let $P$ be  a $(K_{(F,2,mF+r)},F,F-2,mF+r)-$PDA. By Lemma \ref{SA2},
we can obtain two arrays $P_1$ and $P_2$ which are $(K_{(F,2,F)},F,F-2,F)-$PDA and
$(K_{(F,2,(m-1)F+r)},F,F-2,(m-1)F+r)-$PDA from $P$ by
choosing $\frac{F(F-1)}{2}$ columns
and deleting this $\frac{F(F-1)}{2}$ columns, as follow.

\[
P=\left(\begin{array}{c;{2pt/2pt}c}
 P_1    &P_{2}
\end{array}\right)
\]

Thus, $K_{(F,2,(m-1)F+r)}\ge K_{(F,2,mF+r)} - K_{(F,2,F)}$.

Therefore, we have $K_{(F,2,mF+r)} = K_{(F,2,F)} +K_{(F,2,(m-1)F+r)}$.\qed

\section{Conclusion and  problem}

In this paper,  firstly,
we prove two new lower bounds on $S$, when $K,F$ and $Z$ are given.
Secondly,  for some parameters of PDA, we show they are optimal PDAs.
Lastly, for $Z=F-2$, we give a lower bound and
prove that $K_{(F,2,mF+r)} = K_{(F,2,F)} +K_{(F,2,(m-1)F+r)}$ with
$3\le r\le F-3$, $d=$gcd$(F,r)$ and $m>F-r-d$.

For any $m$, if we have  $K_{(F,2,mF+r)} = K_{(F,2,F)} +K_{(F,2,(m-1)F+r)}$ with
$3\le r\le F-3$ and  $d=$gcd$(F,r)$, then we have optimal PDAs for $Z=F-2$.
So we leave them as an open problem.

{\bf Open Problem:} For $Z=F-2$, $3\le r\le F-3$ and  $d=$gcd$(F,r)$, prove that
$K_{(F,2,mF+r)} = K_{(F,2,F)} +K_{(F,2,(m-1)F+r)}$ with $m\le F-r-d$.

\end{document}